\documentclass[twocolumn,epsbox]{jpsj2}
\newcounter{u-large}
\newcounter{u'-large}
\def\runtitle{
Superconductivity in a Two-Orbital Hubbard Model 
}
\def\runauthor{Kazuhiro {\sc Sano} and Yoshiaki {\sc \=Ono} }

\newcommand{\simj}{\stackrel{>}{_\sim}}

\title{%
Superconductivity in a Two-Orbital Hubbard Model with Electron and Hole Fermi Pockets: Application in Iron Oxypnictide Superconductors}

\author{%
Kazuhiro {\sc Sano}\thanks{E-mail address: sano@phen.mie-u.ac.jp} and Yoshiaki {\sc \=Ono}\raisebox{0.5ex}{1,2}  
}

\inst{%
Department of Physics Engineering,  Mie University, Tsu, Mie 514-8507       \\ 
\raisebox{0.5ex}{1}Department of Physics, Niigata University, Ikarashi, Nishi-ku, Niigata 950-2181    \\
\raisebox{0.5ex}{2}Center for Transdisciplinary Research, Niigata University, Ikarashi, Nishi-ku, Niigata 950-2181
}

\recdate{\today}

\abst{%
We investigate the electronic states of a one-dimensional two-orbital Hubbard model with band splitting by the exact diagonalization method. The Luttinger liquid parameter $K_{\rho}$ is calculated  to obtain  superconducting (SC) phase diagram as a function of on-site interactions: the intra- and inter-orbital  Coulomb $U$ and $U'$, the Hund coupling $J$, and the pair transfer $J'$. 
In this model,  electron and hole Fermi pockets are produced when the Fermi level crosses both the upper and lower orbital bands. We find that the system shows two types of SC phases, the SC \Roman{u'-large} for $U>U'$ and the SC \Roman{u-large} for $U<U'$, in the wide parameter region including both weak and strong correlation regimes. Pairing correlation functions indicate that the most dominant pairing for the SC \Roman{u'-large} (SC \Roman{u-large}) is the intersite (on-site) intraorbital spin-singlet with (without) sign reversal of the order parameters between  two Fermi pockets. The result of the SC \Roman{u'-large} is consistent with the sign-reversing $s$-wave pairing that has recently been proposed for iron oxypnictide superconductors. 
}

\kword{%
iron oxypnictide superconductors, two-orbital Hubbard model, pairing symmetry, exact diagonalization
}

\begin{document}
\sloppy
\maketitle

\section{Introduction}

The recent discovery of iron oxypnictide superconductors\cite{Kamihara,Ren,xChen,gChen,Kito} with transition temperatures of up to $T_c \sim 55K$ has stimulated much interest in the relationship between the mechanism of the superconductivity and the orbital degrees of freedom. First-principles calculations have predicted the band structure with hole Fermi pockets around the $\Gamma$ point and  electron Fermi pockets around the $M$ point\cite{Singh,Ishibashi,Nakamura}. By using the weak-coupling approaches based on  multiorbital models, the spin-singlet $s$-wave pairing is predicted, where the order parameter of this pairing changes its sign between  hole and  electron Fermi pockets (sign-reversing $s$-wave pairing)\cite{Kuroki,Mazin,Wang,Nomura,Yanagi}.
This unconventional $s$-wave pairing is expected to emerge owing to the effect of antiferromagnetic spin fluctuations. Since the strong correlation between electrons is considered to play an important role in the superconductivity of iron oxypnictides as well as in that of  high-$T_c$ cuprates, 
 nonperturbative and reliable approaches  are required. 

As a nonperturbative approach, the exact diagonalization (ED) method has been extensively applied in the Hubbard,  $d$-$p$, and  $t$-$J$ models\cite{newref,tJ-Ladder}. Although these models are much simplified and mostly limited to one dimension, it has elucidated some important effects of a strong correlation on  superconductivity.  Using the ED method, we have studied the one-dimensional (1D) two-orbital Hubbard model in the presence of the band splitting $\Delta$. It is  found that the superconducting (SC) phase appears in the vicinity of the partially polarized ferromagnetism when the exchange (Hund's rule) coupling $J$ is larger than its critical value on the order of $\Delta$\cite{Sanohund}. The result suggests that  spin triplet pairing emerges owing to the effect of  ferromagnetic spin fluctuation. In the case of $\Delta=0$, spin triplet superconductivity has also been discussed on the basis of bosonization\cite{Shelton,Shen,Lee} and  numerical\cite{Sakamoto,Shirakawa,Sano-ICM} approaches. Previous works, however, were restricted to the case of a single Fermi surface, and the effects of  electron and hole Fermi pockets on  superconductivity have not been discussed therein. 

In this study,  we investigate the 1D two-orbital Hubbard model with  electron and hole Fermi pockets, where the Fermi level crosses both the upper and lower bands in the presence of a finite band splitting $\Delta$.   Using the ED method, the Luttinger liquid parameter $K_{\rho}$ is calculated  to obtain the SC phase diagram as a function of on-site Coulomb interactions in a wide parameter region including both weak- and strong-correlation regimes.
It would clarify the effects of a strong correlation on  superconductivity in  iron oxypnictides. We also calculate  various pairing correlation functions and discuss a possible pairing symmetry. Although our model is much simplified and limited to one dimension, we expect that the essence of the  superconducting mechanism of iron oxypnictides can be discussed.

\section{Model and Formulation}
We consider the one-dimensional two-orbital Hubbard  model given by the following Hamiltonian:
\begin{eqnarray} 
 H&=&t\sum_{i,m,\sigma}(c_{i,m,\sigma}^{\dagger} c_{i+1,m,\sigma}+h.c.) 
      \nonumber \\
  &+&\frac{\Delta}{2}\sum_{i,\sigma}(n_{i,u,\sigma}-n_{i,l,\sigma}) 
     +U\sum_{i,m}n_{i,m,\uparrow}n_{i,m,\downarrow} 
      \nonumber \\
  &+&U'\sum_{i,\sigma}n_{i,u,\sigma}n_{i,l,-\sigma}
    +(U'-J)\sum_{i,\sigma}n_{i,u,\sigma}n_{i,l,\sigma}
      \nonumber \\
  &-&J\sum_{i} (c_{i,u,\uparrow}^{\dagger} c_{i,u,\downarrow} 
          c_{i,l,\downarrow}^{\dagger} c_{i,l,\uparrow}+h.c.)   
      \nonumber \\
     &-&J'\sum_{i} (c_{i,u,\uparrow}^{\dagger} c_{i,u,\downarrow}^{\dagger}
        c_{i,l,\uparrow} c_{i,l,\downarrow}+h.c.),
\label{hund-Hamil}  
 \end{eqnarray} 
where $c^{\dagger}_{i,m,\sigma}$ stands for the creation operator of an
 electron with a spin $\sigma \ (= \uparrow, \downarrow)$ and an orbital $m \ ( =u,l)$ at site $i$  and 
   $n_{i,m,\sigma}=c_{i,m,\sigma}^{\dagger}c_{i,m,\sigma}$.
Here, $t$ represents the hopping integral between the same orbitals and 
we set $t=1$ in this study.
The interaction parameters $U$, $U'$, $J$, and $J'$ stand for the intra- and inter-orbital direct Coulomb interactions, exchange  (Hund's rule) coupling, and  pair-transfer, respectively.
 $\Delta$ denotes the energy difference between the two atomic orbitals. 
For simplicity, we impose the relation $J=J'$.
%
%
\begin{figure}[t]
\begin{center}
\includegraphics[width=6.5cm]{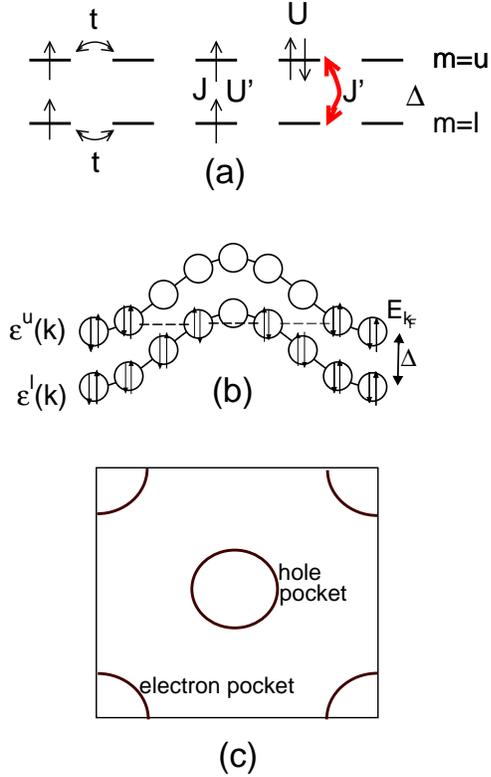}
\end{center}
 \caption{Schematic diagrams of (a) the model Hamiltonian, (b) the band  structure in the noninteracting case, and (c) a corresponding two-dimensional Fermi surface  related to our 1D model. }
 \label{bandmodel}
\end{figure}
%

The model in eq. (\ref{hund-Hamil})  is schematically shown in  Fig. \ref{bandmodel}(a).
In a noninteracting case ($U=U'=J=0$), the Hamiltonian eq. (\ref{hund-Hamil}) yields dispersion relations representing the upper and lower band energies: 
$ \epsilon^{u}(k)=2t \cos(k) + \frac{\Delta}{2}$ and 
$ \epsilon^{l}(k)=2t \cos(k) - \frac{\Delta}{2}$, 
where $k$ is the wave vector. This  band structure  is schematically shown in  Fig. \ref{bandmodel}(b). When the Fermi level $E_{k_{\rm F}}$ crosses  both the upper  and lower bands, the system is  metallic with electron and hole Fermi pockets  corresponding to a characteristic band structure of the FeAs plane in iron oxypnictides, as shown in Fig. \ref{bandmodel}(c). 

We numerically diagonalize the  model Hamiltonian up to 6 sites (12 orbitals) and estimate the Luttinger liquid parameter $K_{\rho}$ from the  ground-state energy of finite-size systems  using the standard Lanczos algorithm.\cite{newref}
To reduce the finite size effect, we impose the boundary condition (periodic or  antiperiodic)  on upper and lower orbitals independently  and chose  both boundary conditions to minimize $|K_{\rho}^0-1|$, where  $K_{\rho}^0$ represents the $K_{\rho}$ of the finite-size system in a noninteracting case. The typical deviation of $K_{\rho}$ from unity becomes about $\sim 0.1$ for a six-site system. 
For simplicity, we will redefine $K_{\rho}$ as a renormalize value calculated using $K_{\rho}/K_{\rho}^0$,  hereafter.

On the basis of the Tomonaga-Luttinger liquid theory \cite{Solyom,Voit,Emery1,Balentz,Fabrizio}, various types of correlation functions are determined by the single parameter $K_\rho$ in the model which is isotropic in spin space. 
For a single-band model with two Fermi points, $\pm k_{F}$, the SC correlation function decays as $\sim r^{-(1+\frac{1}{K_{\rho}})}$, while the CDW and SDW correlation functions decay as $\sim \cos({2k_F r}) r^{-(1+K_{\rho})}$.
Thus, the SC correlation  is dominant for $K_{\rho}>1$, while the CDW or SDW correlation is dominant for $K_{\rho}<1$. 
On the other hand, for a two-band model with four Fermi points, i.e., $\pm k_{F_1}$ and $\pm k_{F_2}$,  low-energy excitations are given by a single gapless charge mode with a gapped spin mode\cite{Balentz,Fabrizio,Emery}. In this case, the SC  and  CDW correlations decay as $\sim r^{-\frac{1}{2K_{\rho}}}$ and $\sim \cos[{2(k_{F_2}-k_{F_1}) r}] r^{-2K_{\rho}}$, respectively, while the SDW correlation decays exponentially. Hence, the SC correlation is dominant for  $K_\rho >0.5$, while the CDW correlation is dominant for $K_\rho <0.5$.
In either case, the SC correlation increases with the exponent $K_\rho $,
 and then  $K_{\rho}$ is regarded as a good  indicator of  superconductivity.\cite{Ladder}
As the noninteracting $K_{\rho}$ is always unity, we assume that the condition of $K_{\rho}>1$ for our model corresponds to   the superconducting state  realized in oxypnictide superconductors.

\section{Phase Diagram}
%
%
%
\begin{figure}[t]
\begin{center}
\includegraphics[width=7.0cm]{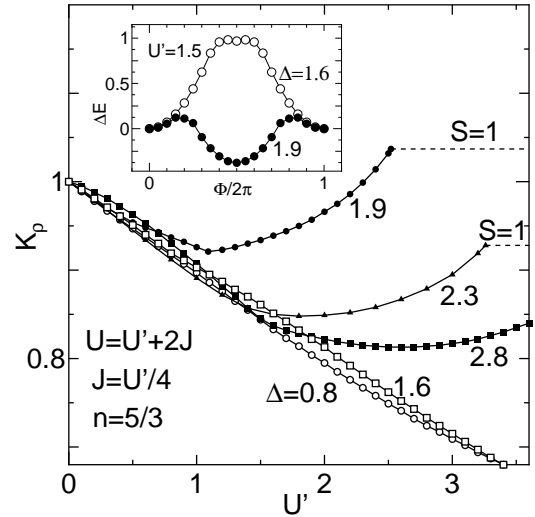}
\end{center}
\caption{$K_{\rho}$ as a function of $U'(=4J)$ for $n=5/3$ (10 electrons/6 sites)  at $\Delta=0.8, 1.6, 1.9, 2.3$, and 2.8. 
The singlet ground state  changes into a partially  polarized ferromagnetic ($S$=1) state   at $U'\simeq$ 2.5, 3.2, and 4.1  for $\Delta=$ 1.9, 2.3, and 2.8, respectively. 
The inset shows  the energy difference $E_0(\phi)-E_0(0)$  as a function of the external flux $\phi$ for $n=2/3$ (6 electrons/9 sites) at $\Delta=1.2$.
}
\label{kr-all}
\end{figure}
%
%
Figure \ref{kr-all} shows   $K_{\rho}$ as a function of  $U'$ for several values of $\Delta$ at an electron density  $n=5/3$ (10 electrons/6 sites), where we set  $J=U'/4$ with  $U=U'+2J$.
When $U'$ increases, $K_{\rho}$  decreases for a small $U'$, while it increases  for a large $U'$ in the case of $\Delta \ge 1.9$, and then  becomes larger than unity for $U'>2.3$ in the case of $\Delta= 1.9$. 
When $J(=U'/4)$ is larger than a certain critical value, the ground state changes into the partially polarized ferromagnetic state with a total spin $S=1$ from the singlet  state with $S=0$. We find that  superconductivity is  most enhanced in the vicinity of the partially polarized ferromagnetic state. To confirm  superconductivity, we calculate  the energy difference of the ground state, $E_0(\phi)-E_0(0)$,  as a function of the external flux $\phi$. As shown in the inset of Fig.  \ref{kr-all},  anomalous flux  quantization is clearly observed for $\Delta=1.9$ while not for $\Delta=0.8$. 
%

%
\begin{figure}[t]
\begin{center}
\includegraphics[width=6.6cm]{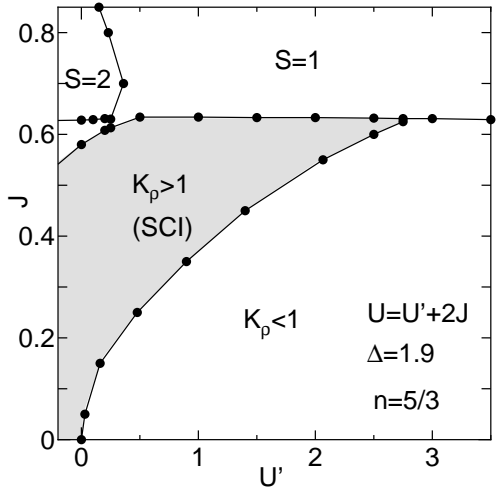}
\end{center}
\caption{Phase diagram of the ground state with  $K_{\rho}$ on the  $U'-J$  parameter plane with $U=U'+2J$ for $n=5/3$ (10 electrons/6 sites)  at $\Delta=1.9$. 
  }
\label{U-Jsouz}
\end{figure}
%
%
\begin{figure}[t]
\begin{center}
\includegraphics[width=6.6cm]{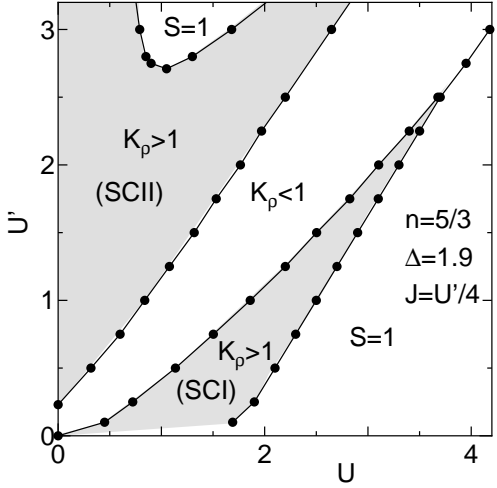}
\end{center}
\caption{Phase diagram of the ground state with  $K_{\rho}$ on the  $U-U'$  parameter plane with  $J=U'/4$ for $n=5/3$ (10 electrons/6 sites)  at $\Delta=1.9$.
  }
\label{U-Upsouz}
\end{figure}

In Fig. \ref{U-Jsouz}, we show the  phase diagram of the ground state  on the $U'-J$ parameter plane  under  the condition of $U=U'+2J$ for $n=5/3$ (10 electrons/6 sites)  at $\Delta=1.9$. 
It contains the  singlet state with  $S=0$  together with partially polarized ferromagnetic states with $S=1$ and  $S=2$.
The singlet state with $K_\rho>1$, where we call it the $SC$ phase,  appears near the partially polarized  ferromagnetic  region at $J \simj U'$. It extends from the attractive  region ($U'<0$) to  the realistic parameter region with $J \sim U'/4>0$, which is expected to correspond to that in the case of iron oxypnictides\cite{Nakamura}. We have confirmed  that  similar phase diagrams are also obtained  for $\Delta=2.3$ and 2.6.

Figure \ref{U-Upsouz} shows the phase diagram of the ground state on the  $U-U'$ plane under  the condition of $J=U'/4$ for $n=5/3$ (10 electrons/6 sites)  at $\Delta=1.9$. 
We observe two types of SC phases with $K_\rho>1$, the SC \Roman{u'-large} for $U>U'$ and the SC \Roman{u-large} for $U<U'$, in the wide parameter region including both weak- and strong-correlation regimes. Note that the SC  \Roman{u'-large}  corresponds to the SC phase shown in  Fig. \ref{U-Jsouz} and belongs to the realistic parameter region mentioned before.

\section{Pairing Correlation}

%
%
\begin{figure}[t]
\begin{center}
\includegraphics[width=6.5cm]{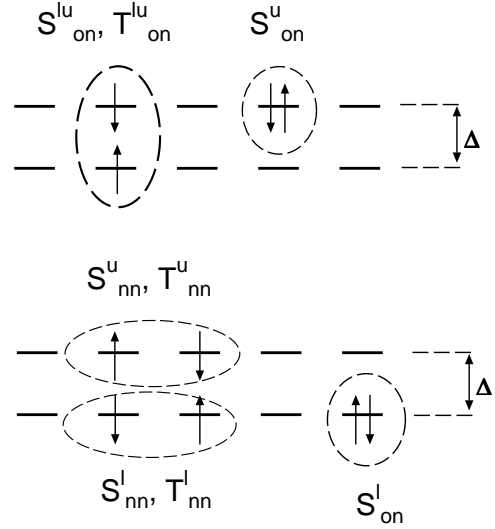}
\end{center}
\caption{Schematic diagrams of various types of  superconducting pairing symmetries, $S^{\rm l}_{\rm on}$, $S^{\rm u}_{\rm on}$, $S^{\rm l}_{\rm nn}$, $S^{\rm u}_{\rm nn}$, and $S^{\rm lu}_{\rm on}$  with spin singlet pairings  and  $T^{\rm l}_{\rm nn}$,  $T^{\rm u}_{\rm nn}$, and $T^{\rm lu}_{\rm on}$ with spin triplet pairings.
  }
\label{hub-pair}
\end{figure}

To examine the nature of these SC phases, we calculate  SC pairing correlation  functions for the various types of pairing symmetries  schematically shown in Fig. \ref{hub-pair}. Explicit forms of the SC pairing correlation  functions $C(r)$  are given by 
\begin{eqnarray}
  S^{\rm l}_{\rm on}(r)&=&\frac1{N}\sum_{i} \langle c^\dagger_{i,l,\uparrow}
 c^\dagger_{i,l,\downarrow} c_{i+r,l,\downarrow} c_{i+r,l,\uparrow}\rangle,
\nonumber \\  
    S^{\rm u}_{\rm on}(r)&=&\frac1{N}\sum_{i} \langle c^\dagger_{i,u,\uparrow} c^\dagger_{i,u,\downarrow}  c_{i+r,u,\downarrow} c_{i+r,u,\uparrow}\rangle,
\nonumber  \\
 S^{\rm l}_{\rm nn}(r)&=&\frac1{2N}\sum_{i}  \langle (c^\dagger_{i,l,\uparrow} c^\dagger_{i+1,l,\downarrow}-c^\dagger_{i,l,\downarrow}c^\dagger_{i+1,l,\uparrow}) 
  \nonumber  \\ &\times& 
(c_{i+r+1\downarrow}c_{i+r,l,\uparrow}-c_{i+r+1,l,\uparrow} c_{i+r,l,\downarrow})\rangle,  
\nonumber \\
      S^{\rm u}_{\rm nn}(r)&=&\frac1{2N}\sum_{i}   \langle (c^\dagger_{i,u,\uparrow} c^\dagger_{i+1,u,\downarrow}-  c^\dagger_{i,u,\downarrow}c^\dagger_{i+1,u,\uparrow})   \nonumber      
\nonumber \\
   &\times& (c_{i+r+1,u,\downarrow}c_{i+r,u,\uparrow}-c_{i+r+1,u,\uparrow} c_{i+r,u,\downarrow})\rangle,  
\nonumber \\ 
 S^{\rm lu}_{\rm on}(r)&=&\frac1{2N_u}\sum_{i}  \langle (c^\dagger_{i,l,\uparrow} c^\dagger_{i,u,\downarrow}-  c^\dagger_{i,l,\downarrow}c^\dagger_{i,u,\uparrow})   \nonumber       
\nonumber \\
 &\times &(c_{i+r,u,\downarrow}c_{i+r,l,\uparrow}-c_{i+r,u,\uparrow}c_{i+r,l,\downarrow})\rangle,  
\nonumber \\ 
  T^{\rm l}_{\rm nn}(r)&=&\frac1{2N_u}\sum_{i}  \langle (c^\dagger_{i,l,\uparrow} c^\dagger_{i+1,l,\downarrow}+ c^\dagger_{i,l,\downarrow}c^\dagger_{i+1,l,\uparrow})   \nonumber   
\nonumber \\
 &\times &(c_{i+r+1\downarrow}c_{i+r,l,\uparrow}+c_{i+r+1,l,\uparrow}c_{i+r,l,\downarrow})\rangle,  
\nonumber \\  
  T^{\rm u}_{\rm nn}(r)&=&\frac1{2N_u}\sum_{i}  \langle (c^\dagger_{i,u,\uparrow} c^\dagger_{i+1,u,\downarrow}+ c^\dagger_{i,u,\downarrow}c^\dagger_{i+1,u,\uparrow})   \nonumber       \\
 &\times &(c_{i+r+1,u,\downarrow}c_{i+r,u,\uparrow}+c_{i+r+1,u,\uparrow} c_{i+r,u,\downarrow})\rangle,  
\nonumber \\ 
 T^{\rm lu}_{\rm on}(r)&=&\frac1{2N_u}\sum_{i} \langle (c^\dagger_{i,l,\uparrow} c^\dagger_{i,u,\downarrow}+c^\dagger_{i,l,\downarrow}c^\dagger_{i,u,\uparrow})   \nonumber       \\
 &\times &(c_{i+r,u,\downarrow}c_{i+r,l,\uparrow}+c_{i+r,u,\uparrow} c_{i+r,l,\downarrow})\rangle,  
\nonumber 
\end{eqnarray}
where   $S^{\rm l}_{\rm on}(r)$, $S^{\rm u}_{\rm on}(r)$, $S^{\rm l}_{\rm nn}(r)$, $S^{\rm u}_{\rm nn}(r)$, and  $S^{\rm lu}_{\rm on}(r)$   denote  the singlet pairing  correlation functions  on the same site in the lower orbital,  on the same site in the upper orbital,  between  nearest-neighbor  sites in the lower orbital,     between  nearest-neighbor sites in the upper orbital, and between the lower and upper orbitals on the same site, respectively. 
Furthermore, $T^{\rm l}_{\rm nn}(r)$,  $T^{\rm u}_{\rm nn}(r)$,  and $T^{\rm lu}_{\rm on}(r)$ are  the triplet pairing   correlation functions between  nearest-neighbor  sites in the lower orbital, between nearest-neighbor  sites in the upper orbital, and  between the lower and upper orbitals on the same site,  respectively.

\begin{figure}[h]
\begin{center}
\includegraphics[width=7.0cm]{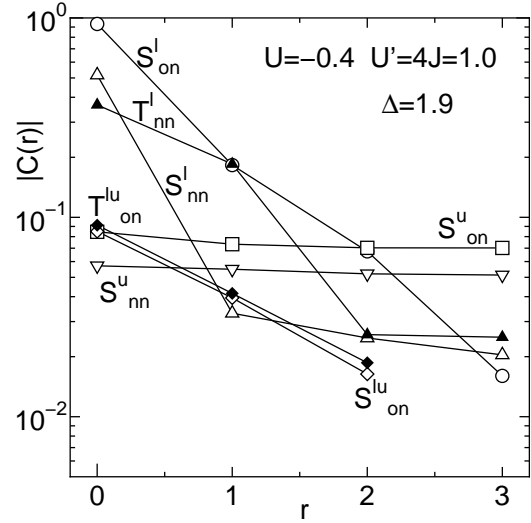}
\end{center}
\caption{Absolute values of various types of  SC pairing correlation  functions  $|C(r)|$  as functions of $r$  for $n$=5/3 (10 electrons/6 sites) at  $\Delta=1.9$, $U'(=4J)=1.0$, and $U=-0.4$, corresponding to the SC \Roman{u-large} phase.}
\label{sc-cor1}
\end{figure}

In Fig. \ref{sc-cor1}, we show the absolute values of various types of  SC pairing correlation  functions  $|C(r)|$  for $n=5/3$ (10 electrons/6 sites) at $\Delta=1.9$,  $U'=4J=1.0$,  and $U=-0.4$. Here, the electronic state of the system belongs to the SC \Roman{u-large}  phase, although the phase diagram for $U<0$ is not explicitly shown in Fig. \ref{U-Upsouz}.  Note that $|T^{\rm u}_{\rm nn}(r)|<10^{-4}$ and $S^{\rm lu}_{\rm on}(r=3)=T^{\rm lu}_{\rm on}(r=3)=0$, which are not shown in Fig. \ref{sc-cor1}. 
We find that $S^{\rm u}_{\rm on}(r)$ and $S^{\rm u}_{\rm nn}(r)$  decay very slowly  as functions of $r$, and  $|S^{\rm u}_{\rm on}(r=3)|$  is the largest among the various $|C(r=3)|$ values. 
Therefore, a relevant  pairing symmetry for the SC \Roman{u-large} phase seems to be the spin singlet pairing in the upper orbital band,  mainly consist of 'on-site' pairing. It is considered that such a pairing in attractive region with $U<0$ is due to the intra-orbital attraction $U$. On the other hand, in repulsive region with $U'>U>0$, the pairing may be due to  charge fluctuation, which is enhanced by a large inter-orbital repulsion $U'$  similarly to that in the case of the $d$-$p$ model in the presence of the inter-orbital repulsion $U_{pd}$\cite{Sano-Udp-all}.

%
%

\begin{figure}[h]
\begin{center}
\includegraphics[width=7.0cm]{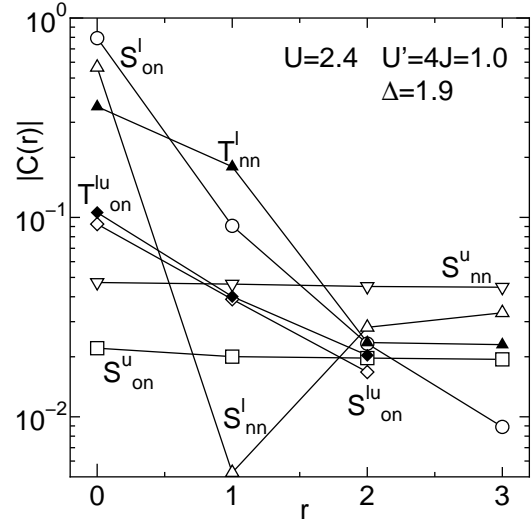}
\end{center}
\caption{Absolute values of various types of  SC pairing correlation  functions  $|C(r)|$  as functions of $r$  for $n$=5/3 (10 electrons/6 sites) at  $\Delta=1.9$, $U'(=4J)=1.0$, and $U=2.4$, corresponding to the SC \Roman{u'-large} phase.}
\label{sc-cor2}
\end{figure}

Next, we discuss the superconductivity in the SC \Roman{u'-large} phase including the realistic parameter region  mentioned before. 
Figure. \ref{sc-cor2} shows  the absolute values of various types of  SC pairing correlation  functions  $|C(r)|$  for $n=5/3$ (10 electrons/6 sites) at $\Delta=1.9$,  $U'=4J=1.0$,  and $U=2.4$, where the system belongs to the SC \Roman{u'-large}  phase, as shown in Fig. \ref{U-Upsouz}.  
Here,  $|T^{\rm u}_{\rm nn}(r)|$,  $|S^{\rm lu}_{\rm on}(r=3)|$, and $|T^{\rm lu}_{\rm on}(r=3)|$  are not shown,  because these correlation functions are very small or zero.
We find that $|S^{\rm u}_{\rm on}(r)|$ is considerably suppressed  compared with $|S^{\rm u}_{\rm nn}(r)|$ in contrast to that in the case of the SC \Roman{u-large} phase. Furthermore,  $|S^{\rm l}_{\rm nn}(r)|$  increases with increasing $r$ except at $r=2$. Therefore, the relevant  pairing symmetry for the SC \Roman{u'-large} phase seems to be an extended spin singlet pairing, and mainly consist of nearest-neighbor site pairing.

%
%

Recently, weak-coupling approaches such as RPA and perturbation expansions have shown that the sign-reversing  $s$-wave ($s_{\pm}$-wave) pairing is realized in  iron oxypnictide superconductors\cite{Kuroki,Mazin,Wang,Nomura,Yanagi}.
 The order parameter of such a pairing is considered to change its sign between  hole and the electron Fermi pockets.  
To compare our result with the result obtained by weak-coupling approaches, we  examine  the SC  pairing correlation function between  the $lower$ and $upper$ orbitals, such as  
\[
  S^{\rm l-u}_{\rm nn}(r)=\frac1{2N}\sum_{i}  <{\Delta^{l}_{\rm nn}(i)}^\dagger \Delta^{u}_{\rm nn}(i+r)>
\]  
with
\[
{\Delta^{m}_{\rm nn}(i)}^\dagger= c^\dagger_{i,m,\uparrow} c^\dagger_{i+1,m,\downarrow}-  c^\dagger_{i,m,\downarrow}c^\dagger_{i+1,m,\uparrow}
\ \   (m=l,u).  
\]
We also define $T^{\rm l-u}_{\rm nn}(r)$ as well as $S^{\rm l-u}_{\rm nn}(r)$ in the above equation.

\begin{figure}[h]
\begin{center}
\includegraphics[width=7.0cm]{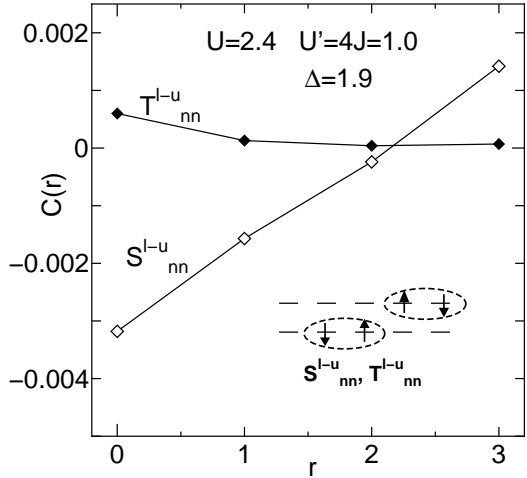}
\end{center}
\caption{Pairing correlation functions  $S^{\rm l-u}_{\rm nn}(r)$ and  $T^{\rm l-u}_{\rm nn}(r)$.
Here, we show  the absolute value of the correlation functions at  $U=2.4$, $\Delta=1.9$, and $U'(=4J)=1.0$ for $n$=5/3 (10 electrons/6 sites).
The inset shows  a schematic diagram of the pairing symmetries, $S^{\rm l-u}_{\rm nn}(r)$  and  $T^{\rm l-u}_{\rm nn}(r)$. 
}
\label{dd-pp}
\end{figure}

When the $s_{\pm}$-wave pairing is dominant, the values of the interorbital SC pairing  correlation function  are expected to be negative, since the Fermi surface of the lower (upper) orbital band in our model corresponds to a hole (electron) Fermi pocket, as shown in Fig. \ref{bandmodel}(b).
 In Fig. \ref{dd-pp}, we show the interorbital pairing  correlation functions  $S^{\rm l-u}_{\rm nn}(r)$  and  $T^{\rm l-u}_{\rm nn}(r)$(see also inset) for the same parameters in Fig. \ref{U-Upsouz} corresponding to the SC \Roman{u'-large}  phase. 
We see that the values of $T^{\rm l-u}_{\rm nn}(r)$  are positive and very small, while those of $S^{\rm l-u}_{\rm nn}(r)$  are negative except at $r=3$ and not so small. 
This result suggests that the relevant pairing symmetry of the SC \Roman{u'-large}  phase is the spin-singlet $s_{\pm}$-wave pairing, which agrees with the result of the weak coupling approaches. 
Therefore, we expect that the $s_{\pm}$-wave pairing proposed on the basis of the result of weak-coupling approaches is realized in the wide parameter region including both weak- and strong-correlation regimes.

%
%
%
%

Finally, we discuss the mechanism of superconductivity in the SC \Roman{u'-large} phase. 
The $s_{\pm}$-wave pairing  is considered to be mediated by the antiferromagnetic fluctuation due to the nesting effect between  electron and hole pockets \cite{Kuroki,Mazin,Wang,Nomura,Yanagi}. At first glance, the SC \Roman{u'-large} phase is located adjacent to  the partial ferromagnetic phase (S=1), and then, the ferromagnetic fluctuation seems to be related to superconductivity. 
To examine the relationship between  spin fluctuation and superconductivity, we also calculate the spin correlation function for  finite-size systems, where a short-range spin correlation is considered to be crucial for the superconductivity.
We obtain the ferromagnetic and antiferromagnetic components of the spin correlation as a function of $U'(=4J)$ for a fixed $U=1.5$ (see also Fig. \ref{U-Upsouz}) and find that  antiferromagnetic (ferromagnetic) correlation increases (decreases) with decreasing $U'$ together with increasing $K_\rho$ (not shown). 
Therefore, we conclude that the antiferromagnetic spin fluctuation is responsible for the $s_{\pm}$-wave pairing in the SC \Roman{u'-large} phase.

%
%
 \section{Summary and Discussion}
We have investigated the superconductivity  of the  one-dimensional two-orbital Hubbard model in the case of electron and hole Fermi pockets  corresponding to a characteristic band structure of iron oxypnictide superconductors. 
To obtain reliable results including those in the strong-correlation regime, we  used  the exact diagonalization method and  calculated the critical exponent $K_{\rho}$ on the basis of  the Luttinger liquid theory. It has been found that the system shows two types of SC phases, the SC \Roman{u'-large} for $U>U'$ and the SC \Roman{u-large} for $U<U'$, in the wide parameter region including both weak- and strong-correlation regimes. 

We have also calculated various types of SC pairing correlation functions in the realistic parameter region of the iron oxypnictides. The result indicates that the most dominant pairing for the SC \Roman{u'-large} phase is the intersite intraorbital spin-singlet with sign reversal of the order parameters between  two Fermi pockets.  The result is consistent with the sign-reversing $s_{\pm}$-wave pairing that has recently been proposed on the basis of the result obtained by  weak-coupling approaches for iron oxypnictide superconductors. This indicates that the $s_{\pm}$-wave pairing  is  realized not only in a weak-correlation regime but also in a strong-correlation regime. 
We have also calculated the spin correlation function and found that antiferromagnetic spin fluctuation is responsible for the $s_{\pm}$-wave pairing in the SC \Roman{u'-large} phase.

As for the SC \Roman{u-large} phase, the most dominant pairing is found to be the on-site intraorbital spin-singlet pairing, which is consistent with the ordinary $s$-wave pairing of BCS superconductors. 
However, the superconducting mechanism of this phase is due to the charge fluctuation enhanced by the interorbital Coulomb interaction and  is different from the conventional BCS superconductivity due to the electron-phonon interaction. 
Although the SC \Roman{u-large} phase seems to be realized only for the unrealistic parameter region in our model, it might be realized  for a realistic parameter region  in  the $d$-$p$ model, which is  closer to  iron oxypnictides\cite{Yanagi,Sano-Udp-all}. 
We will address such a problem by applying the present method in the $d$-$p$ model in the future.



\begin{thebibliography}{99}
\bibitem{Ladder} The noninteracting band structure for the present model is almost equivalent to that for the ladder model, which has   electron and hole Fermi pockets during moderate  band splitting and filling. Results obtained using numerical methods such as the ED and DMRG indicate that the electronic state of the ladder model can be  analyzed well using the weak-coupling theory except in a strong-coupling limit\cite{tJ-Ladder}.
\end{thebibliography}
\end{document}